
\input phyzzx
 \hsize=15.8cm
\vsize=23cm
\voffset=0pt

\newwrite\ffile\global\newcount\figno \global\figno=1
\def\fig{fig.~\the\figno\nfig}
\def\nfig#1{\xdef#1{fig.~\the\figno}%
\writedef{#1\leftbracket fig.\noexpand~\the\figno}%
\ifnum\figno=1\immediate\openout\ffile=figs.tmp\fi\chardef\wfile=
\ffile%
\immediate\write\ffile{\noexpand\medskip\noexpand\item{Fig.\
\the\figno. }
\reflabeL{#1\hskip.55in}\pctsign}\global\advance\figno by1\findarg}

\parindent 25pt
\overfullrule=0pt
\tolerance=10000

\def\half{{\textstyle {1 \over 2}}}

\def\ie{{\it i.e.}}

\nopagenumbers
\baselineskip=14pt

\line{\hfill DAMTP/94-38}
\line{\hfill NSF-ITP-94-52}

\vskip 5cm
\centerline{SUMMING OVER WORLD-SHEET BOUNDARIES}
\vskip 1cm
 \centerline{ Michael B.  Green,}
\centerline{DAMTP, Silver Street, Cambridge CB3 9EW, UK\foot{email:
M.B.Green@amtp.cam.ac.uk}}
\vskip 1cm
\centerline{Joseph Polchinski}
\centerline{ITP, University of California, Santa Barbara, CA 93111, USA\foot
{email: joep@sbitp.ucsb.edu}}
\nopagenumbers
\vskip 4cm
\abstract
The moduli associated with boundaries in a Riemann surface
are parametrized by
the positions and strengths of electric charges.
This suggests a method for summing over orientable Riemann surfaces with
Dirichlet boundary conditions on the embedding coordinates.  A light-cone
parameterization of such boundaries is also discussed.

\vfill\eject
\pagenumbers
\pageno=1
\sequentialequations

The inclusion of  boundaries in the sum over world-sheets that
defines string
perturbation theory alters the properties of  string theory rather
dramatically.  With Neumann boundary conditions on the embedding
coordinates
$X^\mu(\sigma,\tau)$, the resulting theory describes interacting
open and
closed strings -- a boundary representing the trajectory of an
open-string
end-point.  Boundaries with Dirichlet conditions on the embedding coordinates
($X^\mu(\sigma,\tau)=y^\mu$) also have
significant effects---for some discussion see \REF\greenaa{M.B.
Green, {\it Space-Time
duality and Dirichlet
string theory}, Phys.  Lett. {\bf 266B} (1991)
325.}   [\greenaa] and references therein.
 From the target space point of view,
the Dirichlet boundary is simply a point.
This suggests using world-sheet coordinates
in which the boundary is mapped to a point;
what distinguishes this
point is singular behavior of the {intrinsic}
metric $g_{ab}$.  Since the world-sheet now appears to
be topologically trivial, the sum over arbitrary numbers of
Dirichlet boundaries can at least formally be recast as a
world-sheet {field theory}, which is the motivation for this paper.

The strategy to be followed is to reexpress the general metric
for a surface
(of euclidean signature) with an arbitrary number of holes, $\tilde
g$,  as the
sum of  a metric on a surface with no holes, $g$,  and a bilinear
in a vector
field living on the surface,
$$\tilde g_{\alpha\beta} = g_{\alpha\beta} + A_{\alpha}
A_{\beta},\eqn\modmetric$$
 ($\alpha,\beta=1,2$) where $A = -*d\phi$ (in components,
$A_\alpha =
-g_{\alpha\beta} \epsilon^{\beta\gamma} \partial_\gamma \phi$)  is
a 1-form
vector field.
 We will refer to $\tilde g$ in the following as the \lq modified'
metric.  It bears an obvious resemblence to a metric that would
arise in Kaluza--Klein reduction from a three-dimensional theory
(in which $g_{\alpha 3} =A_\alpha$).
The inverse modified metric is given by  $\tilde g^{\alpha\beta}
=
g^{\alpha\beta} - A^\alpha A^\beta(1 + A^\gamma A_\gamma)^{-1}$,
where indices
are here raised by $g^{\alpha\beta}$ (the inverse of $g$).  The
magnetic field
strength due to the vector field is taken to vanish everywhere
except at $B$
arbitrary points,
$$*F\equiv *dA = -d*d\phi = \sqrt g \nabla^2 \phi =  \sum_{i=1}^B
2\pi \alpha_i
\delta^2( w-w_i),\eqn\defina$$
where $\alpha_i$ are arbitrary real parameters (constrained so that
$\sum_{i=1}^B\alpha_i =0$) -- in other words, there are $B$ point
magnetic
vortices on the surface at complex positions $w_i$.
Equivalently, $\phi$ is
the electrostatic potential for $B$ electric charges on the
surface.  The
electric field is given by $E_\alpha = \nabla_\alpha \phi =
g_{\alpha\beta}
\epsilon^{ \beta\gamma}   A_\gamma$.

These charges produce the desired effect.
Close to a charge of strength $\alpha$, located for
convenience at
the origin, the vector potential can be written in complex
coordinates as
$$A_w dw = - i \partial_w \phi dw \sim  -i {\alpha \over 2}{1\over
w}dw =- i
{\alpha\over 2} \left({dr\over r} + i d\theta \right) ,
\eqn\monomet$$
where  $w = r e^{i\theta}$ (and $A_{\bar w} = \bar A_w$).
In polar
coordinates,
$$A_r \equiv {1\over r} (wA_w + \bar w A_{\bar w}) \sim 0, \qquad
A_\theta
\equiv i (w A_w - \bar w A_{\bar w}) \sim  \alpha. \eqn\polvec$$
Therefore, near $r=0$,  $\tilde g_{\theta\theta} \sim \alpha^2$ and
the
modified metric,  $d\tilde s^2 \sim dr^2 + \alpha^2 d\theta^2$,
is that of
the end of a cylinder where  $ r$ is the coordinate along the axis
and the
circumference is $2\pi \alpha$.
The addition of the $A_{\alpha} A_{\beta}$ term in \modmetric\  has
converted
the metric to one appropriate to a surface with extra boundaries.
The
parameters of $A_\alpha$ determine the complex structure on the
surface.   The
complex position of each additional charge and its strength
$\alpha$ are the
three parameters that define the moduli of each additional boundary
so that the
inclusion of $B$ charges of arbitrary strength plausibly defines
metrics that
cover the whole of moduli space for a surface with $B$  boundaries.
The sum
over boundaries can be replaced by a functional integral over
$\phi$  on a
world-sheet with no boundaries.

If $X^\mu$ are scalar \lq matter' fields on the world-sheet (such
as the usual
string space-time coordinates) then $X^\mu(w_i)$ (where $w=w_i$ is
the point at
which a  charge resides on the original world-sheet) becomes the
boundary value
of $X^\mu$ in the description in which the world-sheet metric is
modified and
there is a boundary at $w_i$.   In other words, Dirichlet boundary
conditions
arise naturally, together with the prescription that the boundary
value should
be integrated (since $X^\mu(w)$ is integrated for all values of
$w$).

The hope, therefore, is that there is an effective string theory
including the
vector field that defines the full effect of summing over arbitrary
numbers of
boundaries on a surface with an arbitrary number of handles. Of
course, there
remains the usual problem of summing over the number of handles,
or genus, of
the surface.

\vskip 0.2cm
{\it Two boundaries}

As a simple explicit example consider first the line element of
the constant
curvature sphere in stereographic coordinates,
$$ds^2 =  {1\over (1+r^2)^2}(dr^2 + r^2 d\theta^2)= {1\over (1+w
\bar w)^2} dw
d\bar w , \eqn\polarmet$$
where $w=re^{i\theta}$, $0\le r\le \infty$ and $0\le \theta \le
2\pi$.    The
$r^2$ behaviour of $g_{\theta\theta}$ near the origin indicates the
coordinate
singularity that makes the origin no different from any other
point.  Now
consider the modified metric with  two charges of compensating
strengths
$\alpha$ and $-\alpha$ at the North and South poles (the total
charge vanishes
by Gauss' law since $A_\alpha$ is globally defined).  The modified
spherical
metric is
$$d\tilde s^2 = {1 \over (1+r^2)^2} \left(dr^2 + (r^2 + \alpha^2
(1+r^2)^2)
d\theta^2\right).\eqn \modsphere$$
Equation \modsphere\ defines a metric on a cylinder and can be
rewritten in a
conformally flat manner  by the reparameterization $r\to r'$ defined
by
$$ r' = \int_0^r {dr^{\prime\prime}  \over (r^{\prime\prime 2} +
\alpha^2(1+r^{
\prime\prime 2})^2)^{\half}}  ,\eqn\reparasphere$$
so that
$$d\tilde s^2 =  {r^2 + \alpha^2(1+r^2)^2 \over (1 + r^2)^2} \left(
dr^{\prime
2} + d\theta^2 \right). \eqn\confmet$$
In this form the metric is manifestly conformally equivalent to the
flat metric
on cylinder of circumference $2\pi$ and of length
$$L(\alpha) \equiv r'(r=\infty) - r'(r=0)= \int_0^\infty
{dr^{\prime\prime}
\over (r^{\prime\prime 2} + \alpha^2(1+r^{ \prime\prime
2})^2)^{\half}} ,
\eqn\lengthdef$$
which is finite for finite $\alpha$.  The boundaries of moduli
space are given
by $L(\alpha\to \infty ) \to  0$ and  $L(\alpha\to 0) \sim -\ln
\alpha \to
\infty$.  It is easy to see that the Euler character of the surface
vanishes as
expected from the
Gauss--Bonnet theorem.
More generally,  the two charges may be at asymmetric positions
($w_1$ and
$w_2$,) on the  initial sphere  but this gives a modified metric
in the same
conformal class as the above.

\vskip 0.2cm {\it{Multi-boundary insertions in a disk}}

The connection between the parameters of the charges and the moduli of the
surface is
obscured by the fact that the action of the Mobius group, $SL(2,C)$, on
\modmetric\ is not a
conformal transformation.   Indeed, if it were, the number of
independent real parameters
describing a sphere with $B$ charges would be $3B-7$ (taking the constraint
$\sum_{i=1}^B
\alpha_i$ into account), which is one fewer than the number of moduli of the
surface with $B$ boundaries.

The counting of moduli becomes clearer if surfaces with at least one boundary
component are described by the insertion of charges on a flat disk of unit
radius in
the $w$ plane  (equivalently, the flat upper-half plane) instead of on the
constant curvature sphere.  A surface with $B$
boundaries is
then  described by the insertion of $B-1$  charges  with independent strengths
$\alpha_i$  at complex positions $w_i$ on a disk.  Even though the action of
the Mobius
group ($SL(2,R)$ in this case) on \modmetric\  is again not a conformal
transformation we know that three of these parameters must be redundant since
the surface only has $3B-6$
real moduli, so that the naive counting of parameters now works.   From here on
we will
therefore insert charges on the flat disk, arbitrarily fixing one charge at the
origin and
another on the positive $w$ axis.

 The electrostatic  potential $\phi$ may be
chosen to be the solution of the Laplace equation with sources
\defina\
satisfying $\phi=\phi_B$ on the boundary of the disk, where
$\phi_B$ is an
arbitrary constant.  This solution has the form
$$\phi = \half (\rho + \bar \rho), \eqn\phidefin$$
where
$$\rho \equiv \phi + i \psi = \sum_i \alpha_i \ln\left({w - w_i
\over 1 - w
\bar w_i}\right).\eqn\manmap$$
The vector field is then given by
$$A_w = -{ i\over 2}\sum_i {\alpha_i (1- w_i \bar w_i) \over (w -
w_i)(1 - w
\bar w_i)}, \qquad A_{\bar w} =   {i\over 2}\sum_i {\alpha_i
(1-\bar w_i w_i)
\over (\bar w - \bar w_i)(1 - \bar w w_i)} . \eqn\vwcinrho$$
The normal component of the vector potential vanishes on the
boundary, i.e. $w
A_w + \bar w A_{\bar w} =  A_r =0$ at $|w |=1$.

The expression \manmap\ defines a mapping from the disk to a string
diagram  in
the style of Mandelstam \REF\mandelstama{S.  Mandelstam, {\it
Interacting-string picture
of dual
resonance models}, Nucl. Phys. {\bf  B64 } (1973) 205.}[\mandelstama]  which is
useful for
describing the
modified geometry.  Infinitesimal circles around each of the points
$w_i$ in
the $w$ plane  are mapped into incoming (when $\alpha_i >0$)  or
outgoing
(when $\alpha_i < 0$)  closed strings of widths $2\pi \alpha_i$ in
$\psi$.   If
we choose (for convenience) all $\alpha_i >0$ the diagram
represents a
configuration in which $B-1$ closed strings enter at $\phi =-
\infty$ and join
together in pairs at the turning points,
$\partial \rho / \partial w = 0$,
finally ending at the disk boundary, which is a closed string of
width
$2\pi\alpha_B = - 2\pi \sum_{i=1}^{B-1}\alpha_i$ at $\phi= 0$.
The zeroes of the vector field  (the turning points of this map)
are mapped
into the \lq interaction' points where the cylinders join.  In this
mapping the
$\alpha_i$ are the charges while in [\mandelstama] they were
components of the
light-cone momenta of asymptotic string states.

The metric is diagonal in terms of the variables $\phi$ and $\psi$
defined by
$$d\phi = i A_w d w - i A_{\bar w}d\bar w,   \qquad d\psi =   A_w
d w +
A_{\bar w}d\bar w ,\eqn\amapdef$$
so that
$$\eqalign{d \tilde s^2 = &   dw d\bar w (1+ 2A_w A_{\bar w}) +
dw^2 A_w^2 +
d\bar w^2 A_{\bar w}^2  \cr
                    = & {1\over 4A_wA_{\bar w}}  (d\phi^2 + (1+
4A_wA_{\bar w})
d\psi^2) \cr
                =  &  \left(1 +  {dw \over d \rho}{d\bar w \over
d \bar \rho}
\right) \left(d\phi^2
\left(   {d\rho \over d w}{d\bar \rho \over d \bar w} +
1\right)^{-1}+
d\psi^2\right)
.\cr}\eqn\ametnew$$
 The  metric in the $\rho$ plane in the vicinity of the boundaries
$\rho=\rho(w_i)$ has a simple form.  Since $w -w_i \sim C_i
e^{\rho/\alpha_i}$
(where $C_i =(1 - w_i \bar w_i)  \prod_{i\ne j}\left({w_i - w_j
\over 1-w_i
\bar w_j}\right)^{-\alpha_j/\alpha_i}$) near $w = w_i$ (or $\phi
= -\infty$),
the line element is approximately given by
$$ d\tilde s^2 \sim {d\phi^2\over  \alpha_i^{ 2} } C_i^2 e^{ 2 \phi
/\alpha_i} +
d\psi^2. \eqn\approxmet$$
Changing variables from $\phi $ to $\tau$ on the $i$th string in
the vicinity
of  $\phi = -\infty$, where
$\alpha_i d\tau \sim d\phi C_i     e^{ \phi/\alpha_i}$,
gives
$$\tau \sim  C_i  e^{\phi /\alpha_i} + D_i.\eqn\intvarnew$$
The $B-1$ boundaries at $w_i$ are at finite values of the $\tau$
variable
determined by the parameters of the charges,  whereas they  are at
$\phi = -\infty$.

For example, the two-boundary process is here described by a disk
with a single
charge  at the origin in the $w$ plane.  In this case $\rho\equiv
\phi + i \psi
 = \alpha \ln w $ and  the modified metric becomes
$$d\tilde s^2 = \left( e^{2\phi/\alpha} +   \alpha^2  \right)
           \left(d\phi^2 { e^{2\phi/\alpha} \over
\alpha^2(e^{2\phi/\alpha} +
\alpha^2)}  + d\sigma^2 \right)   \eqn\rhoel$$
(where $0\le \sigma\le 2\pi$ and $-\infty \le \phi \le 0$).
Up to a  (nonsingular) conformal  factor this  metric can be
written as the
flat metric on a cylinder by  changing variables from $\phi$ to
$\tau$,
defined by $\tau= \sinh^{-1} \left( {e^{\phi/\alpha} \over |\alpha|}
\right)$.
The original world-sheet is therefore conformally equivalent to a
cylinder of
circumference $2\pi$ and length $L= \sinh^{-1}(1/|\alpha|)$,
which again becomes infinite in the limit $\alpha\to 0$ and is zero
as
$\alpha\to \infty$.  The measure in $L(\alpha)$ translates into the
measure in
$\alpha$ according to
$$dL = {d\alpha\over |\alpha|}  (\alpha^2 + 1)^{-1/2} \equiv
f(\alpha)
d\alpha.\eqn\lengthsm$$

The modified metric represents a world-sheet with holes of finite
modulus,
rather than punctures.  A world-sheet with a puncture may be
expressed by a
metric of the form
$ d w  d\bar w  / w  \bar w$
near $w =0$ and the world-sheet has an infinite snout located near
$r=0$ where
there is  a curvature singularity.  The integrated curvature, $\int
d^2 w \sqrt
g R = 4\pi$ (where $g\equiv \det g$), appropriate to the insertion
of an
infinitesimal  boundary, or puncture.  By contrast the metric we
are now
considering has the form
$${ (2w \bar w +\alpha^2)d w  d\bar w  \over 2w  \bar w } -
{\alpha^2 d w  d  w
 \over 4 w ^2} - {\alpha^2d\bar  w  d\bar w  \over 4 \bar w
^2}\eqn\holeappmet$$
near $w=0$.
The leading contribution to the curvature from the first term (the
puncture
term) is cancelled by the last two terms.  The world-sheet is
cylindrical and
flat near the boundary with the curvature   located on the interior
of the
sheet.

\vskip 0.2cm {\it  {Covering moduli space.}}

In this section we would like to motivate our expectation that
this parameterization covers moduli space.  We start with the
simplest nontrivial example, $B=3$, which we represent as in the
previous section by the unit
disk with two charges.
Place $\alpha_1$ at the origin and
$\alpha_2$ at $x < 1$ on the positive $x$-axis,
leaving three real
parameters.
It
is easiest to consider first the corner of moduli space corresponding
to three small holes.  This is the same as three long cylinders
(circumference $2\pi$ and length $L_i$)
joined together at one
end and each having a boundary at the other.
This limit corresponds to $x
\to 0$, $\alpha_1/x \to 0$, $\alpha_2/x \to 0$. The lengths are
then $L_1 \sim \ln (x/\alpha_1)$,
$L_2 \sim \ln(x/\alpha_2)$, and $L_3 \sim \ln(1/x)$.
Restricting for now to positive charges gives a 6-fold cover of
moduli space in this region, from permuting the $L_i$.  Equivalently,
this a single cover of Teichmuller space, since the modular group is
just the permutations.   This might have been expected, since
Teichmuller space is topologically $R^3$ and so is
$\alpha_1$-$\alpha_2$-$x$ space, the natural
ranges being $0 < \alpha_1 < \infty$, $0 < \alpha_2 < \infty$, $0 < x <
1$.

Now let us check that the rest of the boundary of Teichmuller space
maps to the boundary of $\alpha_1$-$\alpha_2$-$x$ space.
If, as seems plausible, the mapping has no
folds in the interior, it then gives a single cover everywhere.
There are three types of generic boundary:
{\it (i)} One cylinder degenerating (becoming long).
{\it (ii)}
A strip degenerating, leaving an annulus
 with a thin strip attached across one
boundary.
{\it (iii)}
A strip generating, leaving two annuli
with a thin strip connecting a boundary
from each.
These are obtained in the $\alpha_1$-$\alpha_2$-$x$
parameterization as follows.  In
$1.i$ the boundary $i$ is at the end of the degenerating cylinder ($i =
3$ is the boundary of the disk):
$$\eqalign{
(1.1)\qquad& \alpha_1 \to 0, \qquad \alpha_2,\ x\ {\rm fixed} \quad\cr
(1.2)\qquad& \alpha_2 \to 0, \qquad \alpha_1,\ x\ {\rm fixed} \cr
(1.3)\qquad& x \to 0, \qquad \alpha_1/x,\ \alpha_2/x\ {\rm fixed}.\quad
}$$
In $2.i$, boundary $i$ is at the end of the annulus without the strip:
$$\eqalign{
(2.1)\qquad& \alpha_2 \to \infty ,\qquad x,\ \alpha_1\ {\rm fixed} \qquad\quad
\cr
(2.2)\qquad& \alpha_1 \to \infty,\qquad x,\ \alpha_2\ {\rm fixed} \cr
(2.3)\qquad& x \to 0, \qquad \alpha_1,\ \alpha_2\ {\rm fixed} .
}$$
In $3.i$, boundary $i$ is pinching in the middle:
$$\eqalign{
(3.1)\qquad& x \to 0, \qquad \alpha_1,\ \alpha_2/x\ {\rm fixed} \cr
(3.2)\qquad& x \to 0, \qquad \alpha_2,\ \alpha_1/x\ {\rm fixed} \cr
(3.3)\qquad& x \to 1, \qquad \alpha_1,\ \alpha_2/(1-x)\ {\rm fixed}.
}$$
These limits can be understood by examining the metric in the region of
the degeneration.  Approaching the boundary of
$\alpha_1$-$\alpha_2$-$x$ space in any other way
corresponds to a multiple degeneration.

This extends readily to any $B$, giving a single cover of Teichmuller
space or a $B!$-fold cover of moduli space.
In fact, it all works equally well for charges with any fixed set of
signs.  Allowing the charges to run from $-\infty$ to $\infty$
gives a $2^{B-1} B!$-fold cover; the extra $2^{B-1}$ can be absorbed
into the chemical potential for the charges.

One way of determining whether the parametrization of  metrics in \modmetric\
really
gives a single cover of moduli space might be to relate it to a standard
parametrization,
such as that given by the light-cone gauge.   This is a parametrization in
which  world-sheet
curvature is located at isolated  \lq\lq interaction'' points (turning points
of the boundaries),
which correspond to the moduli  of
the surface.  Given such a relation the measure for integrating over $\alpha_i$
and $w_i$ could be determined in terms of  the integration measure in the
light-cone parametrization.
 Experience with earlier string problems
motivated by
string quantum mechanics in the light-cone frame  [\mandelstama] suggests
that the light-cone parameterization may also be  useful for
visualizing the physical effects of  boundaries.

\vskip 0.2cm {\it  {Generating an arbitrary number of boundaries.}}

The description of moduli space for a world-sheet with an arbitrary
number of
Dirichlet boundaries in terms of charges residing on a disk
suggests a
procedure for expressing the sum over all insertions in terms of a
two-dimensional field theory on the disk.

The integration over metrics in the functional integral that
describes string
theory on a genus zero world-sheet with $B$ boundaries reduces to
integration
over the $(3B-6)$-dimensional moduli space  for $B>2$
(1-dimensional for $B=2$
and 0-dimensional for $B<2 $).  The fiducial metric may be chosen
of the form
$\tilde  g_{\alpha\beta} = \delta_{\alpha\beta} + A_\alpha
A_\beta$,
where $A=-*d\phi$ is the potential for $B-1$ charges with arbitrary
real strengths  $\alpha_i$ at
positions $w_i$ on the unit disk, satisfying $\partial^2 \phi =
\sum_{i=1}^{B-1} \kappa^{B-2} 2\pi \alpha_i \delta(w-w_i)$.
The integration over moduli is given  by integration over
$\alpha_i$ and $w_i$,
so the string functional integral   becomes (for $B>2$)
$$\eqalign{ I_B = & \kappa^{B-2}  \int DX \left(\prod_{i=1}^{B-1}
d^2 w_i
\sqrt{\tilde g(w_i)} d\alpha_i \right) f(\{\alpha_i \}, \{w_i\})
\cr
  & \qquad\qquad \exp\left(-{1\over 2\pi} \int d^2 w \sqrt{ \tilde
g} \tilde
g^{\alpha\beta} \partial_\alpha X^\mu \partial_\beta X_\mu
\right) .
\cr}\eqn\defint$$
Here $\kappa$ is the closed-string coupling constant and the
function $f$ is the
measure on the parameter space of the charges that defines the
measure on the
moduli space of the surface with $B$ boundaries,
and must be determined.

The configuration of charges can be obtained in the classical
theory from an
action of the form,
 $ {i\over 2\pi} \int d^2 w \left(  *F\chi - \sum_{i=1}^{B-1}
\alpha_i \chi(w)
\delta^2(w-w_i)  \right)$.
In this expression   $*F= *dA = - d*d\phi$ and the field $\chi$ is
a Lagrange
multiplier field with equations of motion that enforce  the
condition \defina.

This suggests that \defint\  may be obtained from a functional
integral on the
disk with no holes so that the sum over boundaries would have the
form,
$$\eqalign{ \sum_{B=2}^\infty I_B  & =\sum_{B=2}^\infty
\kappa^{B-2} \int D\chi
 D\phi D X^\mu e^{ {-1\over 2\pi}\int  d^2w \sqrt{\tilde g}
\left(\tilde
g^{\alpha\beta} \partial_\alpha X^\mu \partial_\beta X_\mu   + iF
\chi \right)
} \cr &
 \int   \prod_{i=1}^{B-1}  f(\{\alpha_i \}, \{w_i\})  d^2 w_i
\sqrt{\tilde
g(w_i)} d\alpha_i e^{i\alpha_i \chi(w_i)}.\cr} \eqn\modmag$$
More generally, $A_\alpha$ might be taken to be an independent
vector field in
which case  the condition $A=-*d\phi$ arises from \defina\  in the
gauge
$*dA=0$.

It would be interesting to perform the sum in  \modmag\ explicitly
so that the
series may be reproduced by a functional integral based on an
action that
incorporates new world-sheet fields.  As a schematic illustration
of how this might work consider the action
$$S ={1\over 2\pi } \int  d^2w \left(\sqrt {\tilde g } \tilde
g^{\alpha\beta}
\partial_\alpha
X^\mu  \partial_\beta X_\mu   + i F\chi + \kappa \sqrt { \tilde g
} \int
d\alpha f(\alpha) \cos \alpha\chi \right) \eqn\actfull$$
(where $\tilde g$ is of the form \modmetric).  Expanding the
functional
integral $\kappa^{-1}\int Dg DX D\chi D\phi e^{-S}$ in powers of
$\kappa$
produces an expansion in increasing number of charge insertions and
hence of
Dirichlet boundaries.  Thus, in the $O(\kappa^{-1})$ term the
Lagrange
multiplier $\chi$   enforces the condition $F=0$ that sets
$\phi=0$, leaving
the usual string functional on a disk.  The $O(\kappa^0)$ term
gives rise to
the $B=2$ term in \modmag.   Higher-order terms lead to expressions
like
\modmag\ with a particular form for $f$.
Note that the exponential brings in a factor $1/(B-1)!$, so this
over-covers moduli space by a factor $B$: the outer boundary of the
disk must be treated as distinguished.
Of course, the full measure involves a complicated Fadeev-Popov
determinant; it can take a local form such as \actfull\
when this is represented in terms of ghost fields.
The mapping of the boundary to a point, while very natural
for the $X^\mu$, is less simple for the ghosts;
this may be an obstacle for the idea that we are proposing in this section.

Perhaps the details should not at present be
taken seriously, but this suggests the possibility of expressing the
sum over Dirichlet boundaries in terms of a stringy action that
incorporates some novel features.  The strong similarity of the
modified metric to that arising in Kaluza--Klein compactification
suggests that the sum over boundaries might be naturally expressed
in terms of an action principal on a three-dimensional world-sheet.
Given such a formulation the hope would be that instead of
expanding  in powers of
$\kappa$,  non-perturbative information may be deduced by a mean
field
approximation.

Everything described  explicitly in this paper refers to boundary
corrections
to closed-string tree diagrams, thereby defining a modified
closed-string
theory. The  generalization to world-sheets of higher genus (\ie,
with
handles), is straightforward in principle, thus defining the theory
in which
there is a condensate of Dirichlet boundaries in every order in
closed-string
perturbation theory.   Unlike conventional string  theories this
is known to
have power-behaved fixed-angle cross-sections.  A formulation of
an effective theory, such as that illustrated earlier, should be
of interest in determining how the Dirichlet boundary condensate
affects the massless spectrum.

\vskip 0.5cm
\noindent{\it Acknowledgments}  \hfill\break
M.B.G. is grateful to Orlando Alvarez for
discussions, and
wishes to thank the Miller Institute, Berkeley, for support as
a Visiting
Miller Professor during  September 1993.
The work of J. P. was supported by NSF grants PHY-89-04035 and PHY-91-16964.

\refout
\bye